\documentclass{article}
\usepackage[preprint]{spconfa4}
\usepackage{amsmath}
\usepackage{graphicx}
\usepackage{bm}
\usepackage{amssymb}
\usepackage{booktabs}
\usepackage{multirow}
\usepackage{makecell}
\usepackage{url}
\usepackage{adjustbox}
\usepackage{cite}
\usepackage{siunitx}
\robustify\bfseries
\usepackage[dvipsnames]{xcolor}
\usepackage[colorlinks=true,bookmarks=false]{hyperref}
\hypersetup{allcolors=Black}
\usepackage[capitalize,nameinlink]{cleveref}
\crefname{equation}{}{}
\Crefname{equation}{Equation}{Equations}

\usepackage[hang,flushmargin]{footmisc}

\newcommand{\ul}[1]{\underline{#1}}

\copyrightnotice{\parbox[t]{0.98\textwidth}{\raggedright\footnotesize\noindent
\copyright\ 2026 IEEE. Personal use of this material is permitted.  Permission from IEEE must be obtained for all other uses, in any current or future media, including reprinting/republishing this material for advertising or promotional purposes, creating new collective works, for resale or redistribution to servers or lists, or reuse of any copyrighted component of this work in other works.}}

\title{Downstream-Task-Aware Unified Source Separation}
\name{\begin{tabular}{c}Yoshiki Mitsui$^{1}$, Ryo Aihara$^{1}$, Tatsuhiko Saito$^{1}$, Yoshiki Masuyama$^{2}$,\\
Christoph Boeddeker$^{2}$, Julius Richter$^{2}$, Gordon Wichern$^{2}$, Jonathan Le Roux$^{2}$\end{tabular}}
\address{$^{1}$Mitsubishi Electric Corporation, Kanagawa, Japan \\
$^{2}$Mitsubishi Electric Research Laboratories (MERL), Cambridge, MA, USA}
\begin{document}
\ninept
\maketitle
\begin{abstract}
Task-aware unified source separation (TUSS) enables a single model to handle diverse
separation tasks by conditioning on input prompts.
However, conventional TUSS does not account for downstream task requirements,
such as whether the enhanced speech will be used for human listening or
automatic speech recognition (ASR).
In this paper, we propose a prompt extension framework for TUSS that incorporates
downstream task information into the input prompts and switches the loss function
according to the given prompt during training,
enabling outputs with different signal characteristics at inference time.
Specifically, we introduce an ASR-dedicated prompt paired with a regularized
loss function that reduces speech artifacts to improve ASR robustness,
while the standard prompt is paired with the conventional SNR loss function.
Experiments on the LibriSpeech and JNAS corpora demonstrate that the
proposed joint-training scheme enables a single model to improve ASR performance over noisy input
across a wide range of SNR conditions by selecting the ASR-dedicated prompt,
while maintaining general speech enhancement quality when the standard prompt is used.
\end{abstract}
\section{Introduction}
\label{sec:intro}

Source separation is widely used in various applications,
especially as an acoustic front-end for automatic speech recognition (ASR).
Deep-learning-based single-channel source separation methods~\cite{Hershey2015DC,yu2017pit,luo2019convtasnet,luo2020dprnn,subakan2021sepformer,wang2018spsep}
play an important role in single-channel, low-cost input devices.

Task-aware unified source separation (TUSS)~\cite{Saijo2025TUSS, Paissan2025FasTUSS} 
can be used as a unified model for various purposes, such as speech separation,
speech enhancement, music source separation, and cinematic source separation,
by conditioning the model on different prompts.
However, even when the input signal contains the same source types,
the desirable outputs may differ depending on the user's objective.
For example, speech enhancement for human listening typically aims to minimize residual noise,
whereas for ASR, single-microphone enhancement can degrade performance relative to using
the original noisy input~\cite{Ochiai2024ABSDR}.
To address this issue, ASR-oriented objective functions have been proposed
(e.g., \cite{koizumi2022snri,Li2025MAP}), but they tend to sacrifice perceptual quality for human listening.
Furthermore, directly optimizing for a specific ASR system ties the enhancement model to that backend
and requires a differentiable ASR pipeline during training, increasing both complexity and inflexibility.
Observation adding (OA)~\cite{iwamoto2022oa,wang2024bridging,cui2025reducing} is a simpler
postprocessing alternative that improves ASR performance by blending the enhanced signal
back with the noisy input.
However, because OA is applied only at inference time, it cannot incorporate downstream-task requirements into the training objective. This limitation motivates us to explore a unified training framework that can adapt output characteristics according to downstream-task requirements.

In this paper, we propose a prompt-extension framework for TUSS that incorporates downstream-task information
and uses prompt-dependent loss functions, enabling a single model to generate task-appropriate outputs.
A similar prompt-based control method has been explored for reverberation control~\cite{zhang2023uses},
but our approach readily extends to multiple prompts.
Our experiments demonstrate that this framework enables a single model to realize more flexible source separation across diverse purposes and conditions.

\section{Conventional TUSS}
\label{sec:conventional}

We assume that an observed signal $\bm{x} \in \mathbb{R}^L$ of length $L$ is a mixture $\bm{x} = \sum_{n=1}^N \bm{s}_n$ of $N$ sources $\bm{s}_n \in \mathbb{R}^L$.
The goal of source separation is to estimate the source signals $\bm{s}_n$ from the observed signal $\bm{x}$.
In the case of speech enhancement, we typically consider $N = 2$, where $\bm{s}_1$ is the speech signal 
and $\bm{s}_2$ is the noise signal.

TUSS~\cite{Saijo2025TUSS} is a universal source separation framework capable of 
handling multiple source separation tasks by changing the input prompt of a single model.
For example, we can use the TUSS model for speech enhancement by using the prompts
\texttt{<Speech>} and \texttt{<SFX-mix>}, while the same model can be used for
music source separation with \texttt{<Drums>}, \texttt{<Bass>}, \texttt{<Vocals>} and \texttt{<Other inst.>}.
Here, each prompt (e.g., \texttt{<Speech>}, \texttt{<SFX-mix>}) represents a different 
source type in the input audio.

The TUSS model takes as inputs the mixture signal $\bm{x}$ and prompts.
First, we apply a short-time Fourier transform (STFT) and a band-split encoder~\cite{Luo2023Band,Xu2025TIGER}
to the input signal $\bm{x}$, resulting in a 3D tensor $\bm{Z} \in \mathbb{R}^{D \times T \times F}$,
where $D$ and $F$ are the number of channels and bands of intermediate representation respectively.
Then, the input prompt is embedded via the learnable prompt vector $\bm{p}_n \in \mathbb{R}^{D \times 1 \times 1}$,
which is copied across all $F$ bands and concatenated with $\bm{Z}$ along the $T$ axis,
forming $\bm{Z}' = [\bm{p}_1, ..., \bm{p}_N, \bm{Z}] \in \mathbb{R}^{D \times (N+T) \times F}$.
A cross-prompt module transforms $\bm{Z}'$
into intermediate features $\tilde{\bm{Z}}' = [\tilde{\bm{p}}_{1}, ..., \tilde{\bm{p}}_{N}, \tilde{\bm{Z}}]$,
where $\tilde{\bm{p}}_n \in \mathbb{R}^{D \times 1 \times F}$ denotes each updated prompt embedding,
by jointly processing all input prompts, allowing cross-prompt information exchange.
Prompt-conditioned intermediate features $\tilde{\bm{Z}}_{n}$ are created by replicating $\tilde{\bm{p}}_n$ along the $T$ axis and
applying element-wise multiplication with $\tilde{\bm{Z}}$, i.e., $\tilde{\bm{Z}}_{n} = \tilde{\bm{p}}_n \odot \tilde{\bm{Z}}$.
A target source extraction (TSE) module then refines these into separated source features $\hat{\bm{Z}}_{n}$.
Finally, we obtain the separated source signals $\hat{\bm{s}}_n$ by applying a band-split decoder and inverse STFT
to the separated features $\hat{\bm{Z}}_{n}$.
The cross-prompt and TSE modules consist of TF-Locoformer blocks as described in \cite{Saijo2024TFLoco}.
The TUSS model is optimized using negative signal-to-noise ratio (SNR) loss.

\section{Prompt Extension for Multiple Conditions}
\label{sec:proposed}

\subsection{Overview of Prompt Extension}

We aim to obtain separated signals that reflect additional information,
such as usage scenarios and acoustic conditions.
We propose to introduce new prompts that are trained with downstream-task-specific loss functions,
whereas existing prompts follow the conventional TUSS objective.
This design enables switching between downstream-task-specific outputs at inference time.

\begin{figure}[t]
  \centering
  \includegraphics[width=0.98\columnwidth]{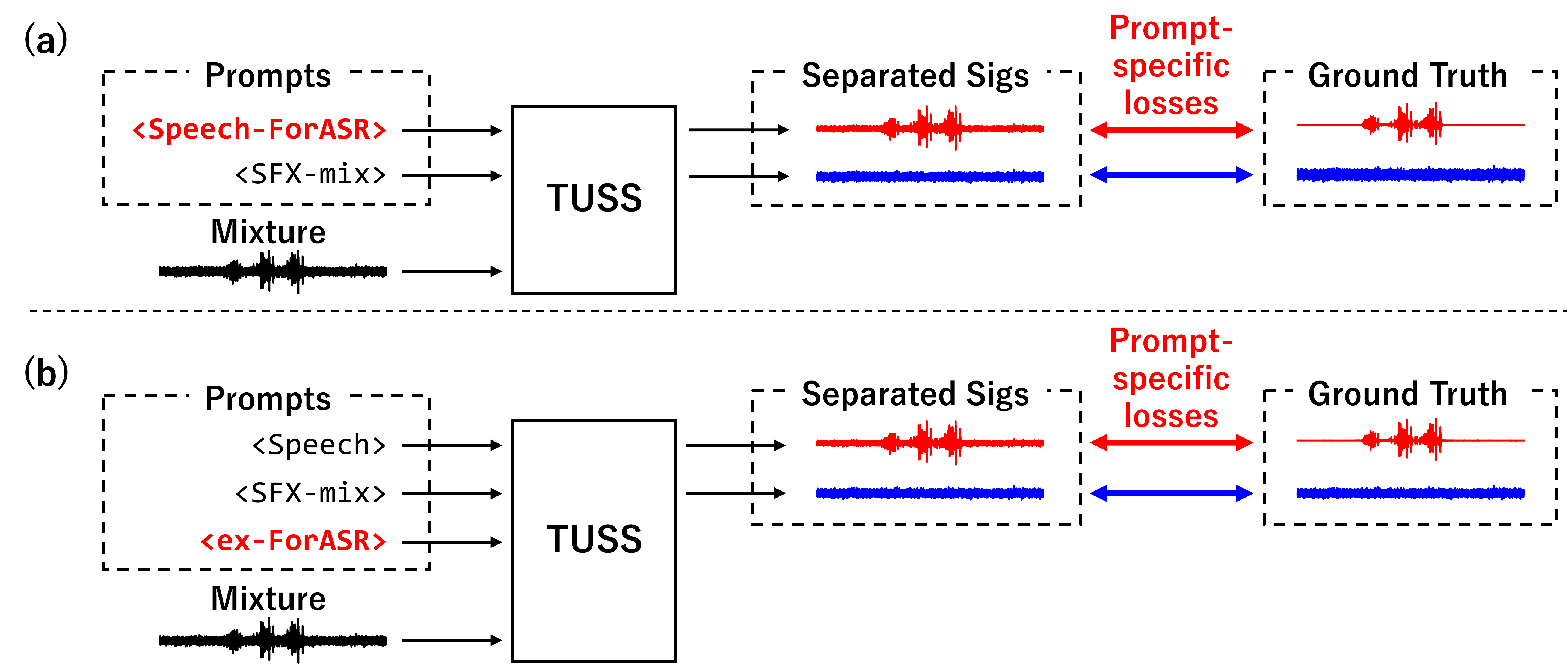}
  \vspace{-.2cm}
  \caption{Overview of prompt extension for task-aware source separation:
  (a) extended prompts incorporating additional downstream task information,
  and (b) additional downstream-task-specific prompts.}
  \vspace{-.4cm}
  \label{fig:concept}
\end{figure}

We introduce two types of downstream-task-specific prompts: \textbf{source-conditioned prompts} are new independently learned
prompt tokens associated with a specific source type and downstream-task (e.g., ASR) requirement,
e.g., \texttt{<Speech-XXXX>};
\textbf{downstream-task-only prompts} encode solely downstream-task information
independent of source identity as an extra (``ex'') token, e.g., \texttt{<ex-XXXX>}.
Downstream-task-only prompts are used only in the cross-prompt module for conditioning and are not passed to the TSE module,
as they do not correspond to explicit output sources.
\Cref{fig:concept} illustrates the proposed framework.
Source-conditioned prompts are generally more suitable for source-specific output requirements,
since each prompt is used to extract a particular source.
By contrast, downstream-task-only prompts are often more effective for global conditions independent of source identity,
such as recording environment and reverberation control~\cite{zhang2023uses}.
Multiple downstream-task-only prompts can also be combined to represent multiple conditions simultaneously.

\subsection{Application to ASR Frontend}

For ASR frontend use, preserving target speech with minimal distortion is
more important than aggressive noise reduction.
In conventional TUSS, the SNR loss does not explicitly enforce these constraints,
which can degrade ASR performance.
MAP-based training~\cite{Li2025MAP} is an ASR-oriented speech-enhancement training method
that adds an L1- or L2-based regularization term between enhanced speech and
a noise-reduced mixture, inspired by maximum a posteriori (MAP) estimation.
The loss function is given by
\begin{equation}
    \mathcal{L}_{\mathrm{L1, R}}(\bm{s}_n, \hat{\bm{s}}_n) = \|\bm{s}_n - \hat{\bm{s}}_n\|_1 + \lambda \| \bm{y}_n - \hat{\bm{s}}_n \|_1,
    \label{eq:l1_reg}
\end{equation}
where $\lambda$ is a regularization weight, and $\bm{y}_n$ denotes a noisy reference signal
whose SNR is increased by $\alpha$~dB relative to the original mixture,
i.e.
\begin{equation}
\bm{y}_n = \bm{s}_n + 10^{-\frac{\alpha}{20}}(\bm{x} - \bm{s}_n).
\end{equation}
The regularization term introduces a trade-off between matching the clean reference $\bm{s}_n$
and the noise-reduced mixture $\bm{y}_n$, which can improve ASR performance by avoiding
overly aggressive noise suppression.
We set the regularization hyperparameters $\lambda$ and $\alpha$
to 1 and 5.0, respectively, which showed good performance in \cite{Li2025MAP}.
However, MAP-based training is designed for a dedicated ASR-oriented model and
does not provide a mechanism to support both general and ASR-oriented enhancement
within a single unified model such as TUSS.
To address this limitation, we introduce new TUSS prompts,
\texttt{<Speech-ForASR>} (source-conditioned) and \texttt{<ex-ForASR>} (downstream-task-only),
which enables a single TUSS model to switch between general and ASR-oriented enhancement
by allowing users to select the appropriate prompt at inference time according to application requirements.
When the \texttt{<Speech-ForASR>} or \texttt{<ex-ForASR>} prompt is used,
we compute the loss function with regularization to prioritize ASR performance.
When downstream-task-specific prompts are not used, we compute the standard loss function to prioritize general waveform reconstruction quality.
Since the original TUSS is trained with an SNR-based objective,
we further propose an SNR-based regularized objective, inspired by MAP-based training:
\begin{equation}
    \mathcal{L}_{\mathrm{SNR, R}}(\bm{s}_n, \hat{\bm{s}}_n) = -\mathrm{SNR}(\bm{s}_n, \hat{\bm{s}}_n) - \lambda \cdot \mathrm{SNR}(\bm{y}_n, \hat{\bm{s}}_n).
    \label{eq:snr_reg}
\end{equation}

\subsection{Application to Reverberation Control}

Input mixtures for source separation are often reverberant, and the desirable amount of reverberation depends on the application: removing reverberation may be preferable for downstream tasks such as ASR, whereas retaining some reverberation may be more desirable for human listening.
To control this behavior, we introduce a new downstream-task-only prompt, \texttt{<ex-KeepRev>}.
A similar control mechanism was introduced in USES~\cite{zhang2023uses} as a memory token.

During model training, we generate reverberant speech by convolving room impulse responses (RIRs) with dry speech and then adding noise.
When \texttt{<ex-KeepRev>} is included, we compute the loss against the reference reverberant speech.
Otherwise, we compute it against the target speech obtained with the earlier 50~ms of the RIR to exclude late reverberation~\cite{kinoshita2013reverb}.
This differs from usual TUSS training, where both mixture and target are without reverberation.

\section{Experiments}
\label{sec:experiments}

\subsection{Experimental Conditions}

To evaluate the effectiveness of the proposed prompt-based TUSS control method,
we conducted experiments on ASR-oriented speech enhancement and reverberation control.

We trained the following TUSS models.
Model A1 was trained with the standard SNR loss using \texttt{<Speech>} and \texttt{<SFX-mix>} as prompts.
Model A2 was trained with the regularized SNR loss in \cref{eq:snr_reg} using the \texttt{<Speech-ForASR>} and \texttt{<SFX-mix>} prompts.
Models A3 and A4 jointly optimize both objectives at 50\% probability:
A3 uses source-conditioned prompts (\texttt{<Speech>} with SNR loss, \texttt{<Speech-ForASR>} with the loss in \cref{eq:snr_reg}),
while A4 applies \cref{eq:snr_reg} when the downstream-task-only prompt \texttt{<ex-ForASR>} is present and the SNR loss otherwise.
In addition to these four settings, we trained three other models (B1--B3),
in which we replaced the SNR losses by their L1 equivalents (e.g., \cref{eq:snr_reg} with \cref{eq:l1_reg}). 
In the results, we distinguish the two output types of A3, A4, and B3 by appending suffixes:
-H denotes outputs for high-quality enhancement,
and -A denotes outputs for ASR-oriented enhancement (using \texttt{<Speech-ForASR>} for A3/B3 and \texttt{<ex-ForASR>} for A4).
For Model A5, -XD and -XR denote outputs without and with the \texttt{<ex-KeepRev>} prompt, respectively,
where X is either H or A, carrying the same meaning as the suffixes defined above.
In all training settings, each mixture contained one speech source and one noise source.
To evaluate joint training for ASR-oriented speech enhancement and reverberation control,
we trained Model A5 with the prompts \texttt{<Speech>}, \texttt{<SFX-mix>}, \texttt{<ex-ForASR>}, and \texttt{<ex-KeepRev>}.
During training, the inclusion of each downstream-task-only prompt (\texttt{<ex-ForASR>} and \texttt{<ex-KeepRev>}) was independently toggled with 50\% probability,
yielding four prompt combinations sampled uniformly at 25\% each.
Models A1--A4 and B1--B3 were trained on anechoic noisy mixtures, whereas Model A5 was trained on an equal mixture of anechoic and reverberant noisy data.

We extracted spectrograms from 16 kHz input mixture signal using STFT
with a 32~ms window length and 8~ms hop length.
Next, we applied a band-split module based on the same frequency partitioning
as in \cite{Xu2025TIGER} (this allocation strategy differs from the original TUSS, which targets 48 kHz input signals). 
All other parameters were set as in the open-sourced TUSS Medium configuration\footnote{\url{https://github.com/merlresearch/unified-source-separation}}.
We used $D = 64$, $K = 4$, $S = 1$, $H = 4$ and $G = 8$ (notations follow \cite{Saijo2024TFLoco}).
We set $(B,E,C)$ to $(4,128,384)$ for cross-prompt modules and to $(2,96,256)$ for TSE modules, where $B$, $E$, and $C$ denote the number of
TF-Locoformer blocks, attention hidden size, and Conv-SwiGLU hidden dimension, respectively.
We trained all models for 150 epochs, with 2000 update steps per epoch and a batch size of 4.
We used the AdamW optimizer with linear warm-up of the learning rate to 0.001 over 10k steps.
If the validation loss did not improve for five epochs, 
we restarted model training from the current best model after halving the learning rate.

For model training and evaluation, we used speech samples from the LibriSpeech~\cite{panayotov2015librispeech} corpus
and noise samples from the ATR Ambient Noise Sound Database II (ATRANS-II)\footnote{\url{https://www.atr-p.com/products/esd.html} (in Japanese)}.
For the speech data, we used 460 hours from the \texttt{train-clean-\allowbreak 100} and \texttt{train-clean-360}
sets for training, 5 hours from the \texttt{dev-clean} set for validation, and 5 hours from the \texttt{test-clean} set for evaluation.
For training the model A5, we made reverberant speech by convolving the anechoic speech with simulated RIRs from the SLR26 database\footnote{\url{https://www.openslr.org/26/}}.
The RIRs were split into training, validation, and test sets in a 7:2:1 ratio.
For the noise data, we split 77 noise types into 54 for training, 16 for validation, and 7 for evaluation.
We removed segments containing clear speech from the noise data using Silero-VAD\footnote{\url{https://github.com/snakers4/silero-vad}}.
At the training and validation stage, we generated anechoic and reverberant noisy mixtures dynamically.
The loudness values in LUFS of speech and noise were uniformly sampled from (-35, -27) and (-33, -25), respectively, for training and validation.
Note that we adopted a lower loudness range for speech sources than noise
because we assumed that the model would be used mainly in severely noisy environments.
We applied speed perturbation to both speech and noise with factors uniformly sampled from (0.9, 1.1).
At the evaluation stage, we generated evaluation mixtures by setting noise loudness to speech +5 LUFS and combined 2,620 speech samples with 7 test noise types (18,340 mixtures in total).
We evaluated speech enhancement quality using SI-SDR~\cite{luo2019convtasnet,leroux2019sdr}, which measures the overall waveform reconstruction accuracy,
UTMOS~\cite{saeki2022utmos}, which evaluates the perceptual audio quality for human listeners,
and word error rate (WER) with the Whisper-turbo model, which evaluates ASR performance.
For baseline comparison, we applied OA with a ratio of 0.4, optimized through preliminary tuning.

To investigate performance trends across a wider SNR range with a different corpus,
we created additional evaluation data for evaluating Japanese ASR performance in noisy conditions.
Five male and five female speakers were chosen from the ASJ Japanese Newspaper Article Sentences (JNAS) Corpus~\cite{itou1999jnas}.
We used the same 7 ATRANS-II noise types reserved for evaluation.
We mixed the speech and noise signals with loudness differences ranging from 5~dB to -15~dB in 5~dB steps.
We refer to the loudness difference as SNR.
In contrast to the English evaluation with WER, we evaluated Japanese ASR performance using character error rate (CER).

\subsection{Results}

\begin{table}[t]
  \centering
  \sisetup{
    reset-text-series = false, 
    text-series-to-math = true, 
    mode=text,
    tight-spacing=true,
    round-mode=places,
    round-precision=1,
    table-format=2.1,
    table-number-alignment=center
}
  \caption{Speech enhancement and ASR performance (SI-SDR [dB], UTMOS, WER [\%]) on LibriSpeech+ATRANS-II.
  SNR\_R and L1\_R are the regularized SNR and L1 losses. Source and Task in the Style column denote source-conditioned and downstream-task-only prompts.
  \texttt{<S>}, \texttt{<N>}, \texttt{<S-A>}, \texttt{<ex-A>} abbreviate \texttt{<Speech>}, \texttt{<SFX-mix>}, \texttt{<Speech-ForASR>}, \texttt{<ex-ForASR>},
  and the bold prompt's output is evaluated.
  $\checkmark$ indicates that OA with a ratio of 0.4 was applied at inference. WER values left/right of the slash are for the 3 easiest/4 hardest noise types.}
  \label{tab:spenh_result}
\begin{adjustbox}{max width=\linewidth}
  \begin{tabular}{llllc|SS[round-precision=2,table-format=1.3]c}
    \toprule
    ID & \makecell[l]{TUSS Obj.} & Style & Prompt & OA & {SI-SDR} & {UTMOS} & {WER} \\
    \midrule
    C & Clean & - & - & - & {-}   & 4.084131 & 3.0 \\
    N & Noisy & - & - & - & -10.5 & 1.41     & 4.5/29.7 \\
    \midrule
    A1 & SNR & - & \textbf{\texttt{<S>}}\texttt{<N>}                            & - &           10.7 &           3.17 &          4.7/21.5 \\ 
    A2 & SNR\_R & - &  \textbf{\texttt{<S-A>}}\texttt{<N>}                      & - &           -6.0 &           1.68 & \textbf{4.2}/20.1 \\
    A3-H & SNR+SNR\_R & Source & \textbf{\texttt{<S>}}\texttt{<N>}              & - & \bfseries 10.8 & \bfseries 3.21 &          4.6/21.1 \\
    A3-A & SNR+SNR\_R & Source & \textbf{\texttt{<S-A>}}\texttt{<N>}            & - &           -5.9 &           1.68 & \textbf{4.2}/19.9 \\
    A4-H & SNR+SNR\_R & Task & \textbf{\texttt{<S>}}\texttt{<N>}                & - &           10.6 &           3.13 &          4.9/22.6 \\
    A4-A & SNR+SNR\_R & Task & \textbf{\texttt{<S>}}\texttt{<N>}\texttt{<ex-A>} & - &           -6.0 &           1.68 & \textbf{4.2}/20.5 \\
    \midrule
    B1 & L1 & - &\textbf{\texttt{<S>}}\texttt{<N>}                    & - & 10.0 & 3.07 & 4.9/25.9 \\
    B2 & L1\_R & - & \textbf{\texttt{<S-A>}}\texttt{<N>}              & - & -3.5 & 1.84 & 4.3/21.0 \\
    B3-H & L1+L1\_R & Source & \textbf{\texttt{<S>}}\texttt{<N>}      & - & 10.0 & 3.07 & 4.9/26.3 \\
    B3-A & L1+L1\_R & Source & \textbf{\texttt{<S-A>}}\texttt{<N>}    & - & -3.5 & 1.84 & 4.3/21.0 \\
    \midrule
    A1' & SNR & - & \textbf{\texttt{<S>}}\texttt{<N>} & $\checkmark$ & -3.4 & 1.86 & \textbf{4.2}/\textbf{18.9} \\
    B1' & L1 & - &\textbf{\texttt{<S>}}\texttt{<N>}   & $\checkmark$ & -3.4 & 1.84 &          4.3/21.3 \\
    \bottomrule
  \end{tabular}%
\end{adjustbox}
\vspace{-.4cm}
\end{table}

\begin{table*}[t]
  \centering
  \caption{Character error rates (CER) [\%] on JNAS+ATRANS-II.
  CER values left/right of the slash are for the 3 easiest and 4 hardest noise types, respectively.
  Underlined scores indicate improvements over A1 (for SNR-loss variants) or B1 (for L1-loss variants).
  Boldfaced scores indicate the best performance among all conditions except Clean.}
  \label{tab:asr_result_cer}
  \begin{adjustbox}{max width=0.8\linewidth}
    \begin{tabular}{llllc|ccccc}
      \toprule
      ID    & TUSS Objective & Style & Prompt                                           & OA           & +5dB                                &  0dB                                  &  -5dB              &  -10dB                         &  -15dB  \\
      \midrule
      C     &  Clean      & -       & -                                                 & -            & \multicolumn{5}{c}{5.1} \\
      N     &  Noisy      & -       & -                                                 & -            &      \textbf{5.3}/7.1               &               5.6/12.5                &      7.2/40.3      &      13.1/263.4                &               33.2/423.1 \\
      \midrule
      A1    &  SNR        & -       & \textbf{\texttt{<S>}}\texttt{<N>}                 & -            &               5.5/7.1               &               5.8/11.5                &      7.0/34.8      &      11.7/170.3                &               29.2/233.2 \\
      A2    &  SNR\_R     & -       & \textbf{\texttt{<S-A>}}\texttt{<N>}               & -            & \textbf{\ul{5.3}}/\textbf{\ul{6.3}} & \textbf{\ul{5.5}}/\phantom{0}\ul{9.7} & \textbf{\ul{6.2}}/\ul{24.5} &  \phantom{0}\ul{9.5}/\ul{146.2} &          \ul{23.2}/384.7 \\
      A3-A  &  SNR+SNR\_R & Source  & \textbf{\texttt{<S-A>}}\texttt{<N>}               & -            & \textbf{\ul{5.3}}/\textbf{\ul{6.3}} & \textbf{\ul{5.5}}/\phantom{0}\ul{9.5} &          \ul{6.4}/\ul{23.4} &            \ul{10.0}/\ul{137.6} & \textbf{\ul{21.5}}/378.5 \\
      A4-A  & SNR+SNR\_R  & Task    & \textbf{\texttt{<S>}}\texttt{<N>}\texttt{<ex-A>}  & -            & \textbf{\ul{5.3}}/\textbf{\ul{6.3}} & \textbf{\ul{5.5}}/\phantom{0}\ul{9.7} & \textbf{\ul{6.2}}/\ul{24.6} &  \phantom{0}\ul{9.4}/\ul{145.9} &          \ul{22.1}/391.3 \\
      \midrule
      B1    &  L1         & -       & \textbf{\texttt{<S>}}\texttt{<N>}                 & -            &               5.5/7.7               &               5.9/12.6                 &       7.1/35.2      &      12.2/124.1           &      31.2/\textbf{146.8}   \\

      B2    &  L1\_R      & -       & \textbf{\texttt{<S-A>}}\texttt{<N>}               & -            & \textbf{\ul{5.3}}/\ul{6.4}          & \textbf{\ul{5.5}}/\phantom{0}\ul{9.8}  &  \ul{6.3}/\ul{25.7} & \textbf{\phantom{0}\ul{9.1}}/138.2 & \ul{24.2}/353.6   \\
      B3-A  &  L1+L1\_R   & Source  & \textbf{\texttt{<S-A>}}\texttt{<N>}               & -            & \textbf{\ul{5.3}}/\ul{6.4}          & \textbf{\ul{5.5}}/\phantom{0}\ul{9.9}  &  \ul{6.3}/\ul{25.0} & \phantom{0}\ul{9.3}/133.9 & \ul{23.0}/351.5   \\
      \midrule
      A1'   &  SNR        & -       & \textbf{\texttt{<S>}}\texttt{<N>}                 & $\checkmark$ & \textbf{\ul{5.3}}/\textbf{\ul{6.3}} & \textbf{\ul{5.5}}/\phantom{0}\textbf{\ul{9.4}} & \textbf{\ul{6.2}}/\textbf{\ul{23.3}} & \phantom{0}\ul{9.2}/\textbf{\ul{117.5}} & \ul{22.9}/359.7 \\
      B1'   &  L1         & -       & \textbf{\texttt{<S>}}\texttt{<N>}                 & $\checkmark$ & \textbf{\ul{5.3}}/\ul{6.4}          & \textbf{\ul{5.5}}/\phantom{0}\ul{9.9}          &  \ul{6.3}/\ul{26.8}                  & \phantom{0}\ul{9.3}/135.1               & \ul{23.9}/354.8 \\
      \bottomrule
    \end{tabular}%
  \end{adjustbox}
  \vspace{-.4cm}
\end{table*}

\begin{table}[t]
  \centering
    \sisetup{
    reset-text-series = false, 
    text-series-to-math = true, 
    mode=text,
    tight-spacing=true,
    round-mode=places,
    round-precision=1,
    table-format=2.1,
    table-number-alignment=center
}
  \caption{Anechoic and reverberant speech enhancement and ASR performance (SI-SDR [dB], WER [\%]) on LibriSpeech+\allowbreak SLR26+\allowbreak ATRANS-II dataset.
  \texttt{<ex-R>} denotes \texttt{<ex-KeepRev>}.
  SI-SDR$_\text{A}$ evaluates outputs from anechoic noisy mixtures using anechoic speech references,
  whereas SI-SDR$_\text{NR}$ and SI-SDR$_\text{R}$ evaluate outputs from reverberant noisy mixtures
  using early-RIR convolved speech and full-RIR convolved speech as references, respectively.
  WERs are computed from anechoic noisy mixtures as in \Cref{tab:spenh_result}.}
  \label{tab:rev_spenh_result}
\begin{adjustbox}{max width=\linewidth}
  \begin{tabular}{lll|Sc|SS}
    \toprule
    ID & \makecell[l]{Training \\ Data} & Prompt & {\!SI-SDR$_\text{A}$\!} & WER & {\!SI-SDR$_\text{NR}$\!} & {\!SI-SDR$_\text{R}$\!} \\
    \midrule
    N & (Noisy) & - & -10.5 & 4.5/29.7 & -11.6 & -11.0 \\
    \cmidrule(lr){1-7}
    A1 & \!Anechoic\! & \textbf{\texttt{<S>}}\texttt{<N>} & \bfseries 10.7 & 4.7/\textbf{21.5} & 5.6 & 6.1 \\
    \cmidrule(lr){1-7}
    A5-HD & Both & \textbf{\texttt{<S>}}\texttt{<N>} & 10.2 & 4.8/24.1 & \bfseries 7.5 & 6.8 \\
    A5-HR & Both & \textbf{\texttt{<S>}}\texttt{<N>}\texttt{<ex-R>} & 10.2 & 4.9/24.0 & 6.6 & \bfseries 7.9 \\
    A5-AD & Both & \textbf{\texttt{<S>}}\texttt{<N>}\texttt{<ex-A>} & -6.0 &  \textbf{4.2}/\textbf{21.5} & -7.4 & -7.0 \\
    A5-AR & Both & \textbf{\texttt{<S>}}\texttt{<N>}\texttt{<ex-A>}\texttt{<ex-R>} & -6.0 & \textbf{4.2}/21.7 & -7.3 & -6.8 \\
    \bottomrule
  \end{tabular}%
\end{adjustbox}
\vspace{-.4cm}
\end{table}

\Cref{tab:spenh_result} shows the speech enhancement and ASR performance.
Comparing A1 and A3-H/A4-H,
for the output signal from the high-quality enhancement branch,
the model achieves almost the same SI-SDR and UTMOS as A1.
On the other hand, comparing A2 and A3-A/A4-A,
the output signal from the ASR-oriented enhancement branch preserves noticeable noise,
leading to lower SI-SDR and UTMOS while prioritizing ASR robustness through the regularization term,
which intentionally balances residual noise against speech distortion.
Overall, changing the input prompt and loss function enables the model to generate signals with
different characteristics, while the jointly trained models A3-H/A and A4-H/A maintain nearly the
same output quality as the single-objective models A1 and A2.
The results for B1--B3, in which the loss function was changed from SNR to L1,
showed similar trends to those for A1--A3 that used the SNR loss,
but the overall performance was lower.
In the ASR evaluation, we report WER separately for the three easiest (running car, running lorry, airport)
and four hardest (station, supermarket, rugby stadium, train) noise types,
because averaging across all seven can obscure meaningful differences.
For A2, despite significantly lower SI-SDR and UTMOS values, the WER is lower than that of A1.
This suggests that regularization using the SNR loss against a noise-suppressed reference
contributes to improved ASR performance.
Furthermore, jointly optimized models A3-H/A4-H and A3-A/A4-A show almost identical performances to the single-objective models A1 and A2, respectively.
This implies that the proposed joint-training scheme enables a single speech-enhancement model to support
both objectives without performance degradation,
as the cross-prompt module is designed to flexibly process multiple prompts simultaneously.
However, compared with A1' (obtained by applying OA), the WERs of A2, A3-A, and A4-A remain higher, suggesting that OA may still better preserve ASR-relevant cues.
Similar trends were observed when replacing the SNR-based objective with an L1-based objective.

\Cref{tab:asr_result_cer} presents the Japanese ASR performance in terms of CER.
Overall, the Japanese CER results follow the same trends as the LibriSpeech-based English evaluation,
suggesting that the findings generalize across languages and corpora.
For the three easiest noise types, comparing N (no speech enhancement) and A1 (standard TUSS-based enhancement),
we observe that ASR performance degrades relative to the no-enhancement baseline, particularly under relatively favorable conditions (SNR $>0$ dB).
This confirms that speech enhancement improves perceptual quality but does not necessarily improve ASR performance.
Comparing the CERs for the three easiest and four hardest noise types,
the CERs for the hardest noises are roughly consistent with those for the easiest noises when the SNR is shifted by approximately 10 dB.
An exception was observed at $-15$ dB SNR for the harder noise types: B1
achieved the best CERs,
while models with the regularized objective such as A3-A performed worse.
Inspection of the ASR outputs revealed frequent repetition errors in the latter models but not in B1,
likely because the L1-based loss suppresses more speech-like artifacts, thereby reducing such errors.

\begin{figure}[t]
  \centering
  \includegraphics[width=0.98\columnwidth]{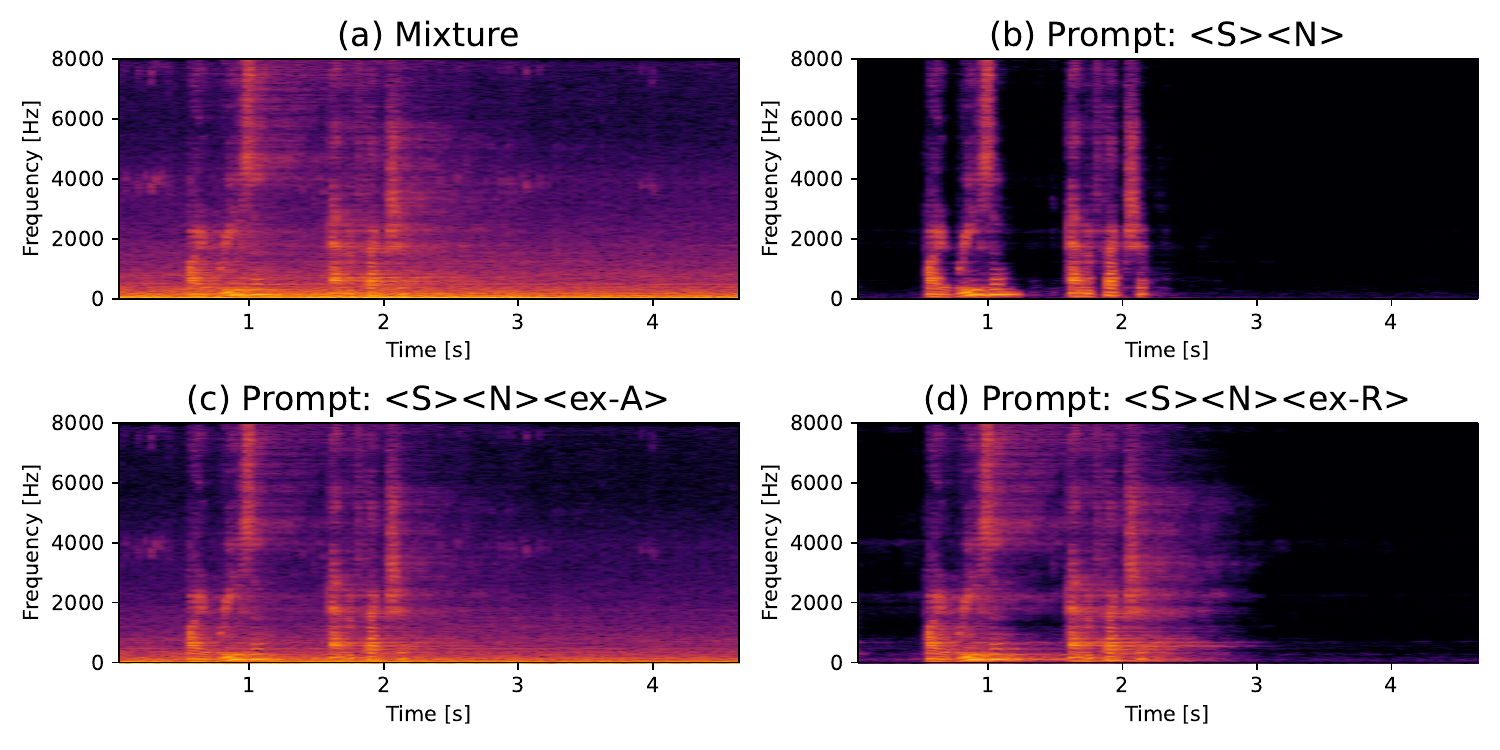}
  \vspace{-.5cm}
  \caption{Examples of output signals generated by Model A5 for different input prompts.}
  \vspace{-.5cm}
  \label{fig:sig_sample}
\end{figure}

\Cref{tab:rev_spenh_result} shows the anechoic and reverberant speech enhancement performance.
Here, SI-SDR$_\text{A}$, SI-SDR$_\text{NR}$, and SI-SDR$_\text{R}$ denote the three evaluation settings
defined in \cref{tab:rev_spenh_result}.
The SI-SDR$_\text{A}$ values of A5-HD and A5-HR are slightly lower, but still nearly comparable,
to that of A1, which was trained only on anechoic data.
A5-HD performs best in SI-SDR$_\text{NR}$, whereas A5-HR performs best in SI-SDR$_\text{R}$,
showing that reverberation control can be achieved by introducing the \texttt{<ex-KeepRev>} prompt.
Among the listed conditions, A5-AD shows better WER performance,
demonstrating that the regularization successfully maintains the model's ASR-oriented characteristics.
Together, these results suggest that the proposed framework can jointly support
ASR-oriented speech enhancement and reverberation control within a single model.
Comparing A5-AD and A5-AR, SI-SDR$_\text{A}$ and SI-SDR$_\text{NR}$ remain nearly unchanged.
This may be because the \texttt{<ex-ForASR>} prompt also has the effect of retaining residual reverberation, similar to the \texttt{<ex-KeepRev>} prompt.
However, SI-SDR$_\text{R}$ shows a modest improvement, suggesting additive benefits from combining the two downstream-task-only prompts.
\Cref{fig:sig_sample} illustrates examples of output signals obtained with Model A5.
These examples show that the model can generate signals with different characteristics depending on the input prompt.

\section{Conclusion}
\label{sec:conclusion}

We proposed a prompt extension method for TUSS that incorporates additional information
into the prompts and enables the use of different loss functions during model training,
thereby generating outputs with different signal characteristics.
Our experiments demonstrated that the TUSS model trained with extended prompts achieves
comparable speech enhancement performance while improving ASR robustness, validating
the effectiveness of the proposed method.
Furthermore, we demonstrated that the proposed framework can be extended to handle
multiple downstream tasks simultaneously, such as ASR-oriented speech enhancement and reverberation control. 
However, our regularized objective still underperforms the baseline OA method,
suggesting that future work should investigate better loss functions for ASR-oriented speech enhancement.

\bibliographystyle{IEEEbib}
\bibliography{refs}

\end{document}